\documentclass[prd,amsfonts,amssymb,
amsmath,showkeys,showpacs]{revtex4}

\date{\today}
\usepackage{graphicx}

\begin{document}

\newcommand{\pslash}[1]{#1\llap{\sl/}}
\newcommand{\kslash}[1]{\rlap{\sl/}#1}
\newcommand{\qq}{$^\sharp$ }
% item that is questionable.
\newcommand{\rmb}{{\color{red}$^\bigstar$} }
% item that is for simple reading.
\newcommand{\fp}{{\color{blue}$^\heartsuit$} }
% item that contains proof usually for reference.
\newcommand{\qa}{$^\mho$ }
% item that contains answer to the question.
\newcommand{\rp}{{\color{magenta}$^\Re$} }
% reference point.
\newcommand{\lab}[1]{\hypertarget{lb:#1}}
% to create labels.
\newcommand{\iref}[2]{\hyperlink{lb:#1}{\textit{$^\spadesuit$#2}}}
% reference to other part in this notes.
\newcommand{\emp}[1]{{\bf\color{red} #1}}
% Emphasis
\newcommand{\eml}[1]{{\color{blue} #1}}
% Quick Guide.
\newcommand{\kw}[1]{\emph{#1}}
% Key word.
\newcommand{\sos}[1]{{\large \textbf{#1}}}
% Subsection
\newcommand{\soso}[1]{{\Large \begin{center} \textbf{#1} \end{center}}}
% Subsection ver 2. Maybe used to do more. One can change the "\[" part to use other like "\center" to recover the "\[" for equation number.
\newcommand{\sossub}[1]{{\large\[\textbf{#1}\]}}
% SubSubsection
\newcommand{\et}[1]{\begin{flushleft} \color{blue}\textbf{#1} \end{flushleft}}
% \newcommand{\et}[1]{{\color{blue}\textbf{#1}}\\}
% Entry
\newcommand{\rf}[1]{{\color{blue}[\textit{#1}]}}
% Reference
\newcommand{\el}[1]{\label{#1}}
% Equation labeling
%\newcommand{\er}[1]{\eqref{#1}}
\newcommand{\er}[1]{Eq.~\eqref{#1}}
% Equation Reference
\newcommand{\df}[1]{\textbf{#1}}
% Temporarily replace \textbf
\newcommand{\mdf}[1]{\pmb{#1}}
% Use for vectors etc.
\newcommand{\n}[1]{$#1$}
% Use for numbers etc.
\newcommand{\cjktext}[1]{\begin{CJK}{GB}{gbsn} #1 \end{CJK}} 
% Language support
\newcommand{\fals}[1]{$^\times$ #1}
% wrong statement
\newcommand{\new}{{\color{red}$^{NEW}$ }}
% update
\newcommand{\de}[1]{{\color{green}\underline{#1}}}
\newcommand{\ke}{\rangle}
\newcommand{\br}{\langle}
\newcommand{\lb}{\left(}
\newcommand{\rb}{\right)}
\newcommand{\blb}{\Big(}
\newcommand{\brb}{\Big)}
\newcommand{\nn}{\nonumber \\}
\newcommand{\p}{\partial}
\newcommand{\pd}[1]{\frac {\partial} {\partial #1}}
\newcommand{\cc}{$>$}
% ##### ###### ###### ######
\newcommand{\ba}{\begin{eqnarray}}
\newcommand{\ea}{\end{eqnarray}}
\newcommand{\be}{\begin{equation}}
\newcommand{\ee}{\end{equation}}
\newcommand{\bay}[1]{\left(\begin{array}{#1}}
\newcommand{\eay}{\end{array}\right)}
\newcommand{\eg}{\textit{e.g.} }
\newcommand{\ie}{\textit{i.e.} }
\newcommand{\iv}[1]{{#1}^{-1}}
\newcommand{\st}[1]{|#1\ke}
% ##### ###### ###### ######
% Greek Letters
\newcommand{\xa}{\alpha}
\newcommand{\xA}{\Alpha}
\newcommand{\xb}{\beta}
\newcommand{\xB}{\Beta}
\newcommand{\xd}{\delta}
\newcommand{\xD}{\Delta}
\newcommand{\xe}{\epsilon}
\newcommand{\xE}{\Epsilon}
\newcommand{\xve}{\varepsilon}
\newcommand{\xg}{\gamma}
\newcommand{\xG}{\Gamma}
\newcommand{\xk}{\kappa}
\newcommand{\xK}{\Kappa}
\newcommand{\xl}{\lambda}
\newcommand{\xL}{\Lambda}
\newcommand{\xo}{\omega}
\newcommand{\xO}{\Omega}
\newcommand{\xs}{\sigma}
\newcommand{\xS}{\Sigma}
\newcommand{\xt}{\theta}
\newcommand{\xT}{\Theta}
% ##### ###### ###### ######
\def \Tr {{\rm Tr}}
\def\CA{{\cal A}}
\def\CC{{\cal C}}
\def\CH{{\cal H}}
\def\CP{{\cal P}}
\def\CL{{\cal L}}
\def\CN{{\cal N}}
\def\CV{{\cal V}}
\def\CR{{\cal R}}
\def\CS{{\cal S}}
\def\CW{{\cal W}}
\def\sN{\mathscr{N}}
\def\ad#1#2{{\delta\over\delta\sigma^{#1}(#2)}}

\author{Xing Huang}
\author{Leonard Parker}
\affiliation{Physics Department, University of Wisconsin-Milwaukee,
P.O.Box 413, Milwaukee, Wisconsin USA 53201}

\title{Hermiticity of the Dirac Hamiltonian in Curved Spacetime}
\date{November 13, 2008}

\begin{abstract}
In previous work on the quantum mechanics of an atom freely falling in a general curved background spacetime, the metric was taken to be sufficiently slowly varying on time scales relevant to atomic transitions that time derivatives of the metric in the vicinity of the atom could be neglected.  However, when the time-dependence of the metric cannot be neglected, it was shown that the Hamiltonian used there was not Hermitian with respect to the conserved scalar product.  This Hamiltonian was obtained directly from the Dirac equation in curved spacetime. This raises the paradox of how it is possible for this Hamiltonian to be non-hermitian. Here, we 
show that this non-hermiticity results from a time dependence of the position eigenstates that enter into the Schr{\"o}dinger wave function, and we write the expression for the Hamiltonian that is Hermitian
for a general metric when the time-dependence of the metric is not neglected.
\end{abstract}
\newpage

\maketitle

\section{Introduction}
In  \cite{Parker:1980kw} and \cite{Parker:1980prl}, a one-electron atom was investigated as a probe of the curvature of a general spacetime.  If the curvature near the atom is sufficiently strong, then the spectrum of the atom can reveal properties of the Riemann tensor at the position of the atom. To calculate the shifts in the energy eigenvalues of the atom by means of perturbation theory, a conserved scalar product suitable to the Dirac equation in a general curved spacetime was defined in \cite{Parker:1980kw}.  This scalar product was based on a generally covariant current introduced by 
Bargmann \cite{Bargmann1932} in developing the
theory of the curved-spacetime Dirac equation obtained 
by Schr{\"o}dinger \cite{Schrodinger1932}.
The Hamiltonian for the one-electron atom was found 
in \cite{Parker:1980kw} directly from the curved-spacetime Dirac equation.  Assuming that the rate of change of the
spacetime curvature in the vicinity of the atom was negligible relative to the transition rates associated with the atom, that Hamiltonian was found to be hermitian with respect to the conserved scalar product,
and the shifts in the energy eigenvalues were obtained in 
terms of the Riemann tensor at the position of the atom.

In \cite{Parker:1980kw}, it was also found that if the time dependence of the metric can not be neglected, then the expression for the Hamiltonian coming directly from the curved-spacetime Dirac equation will violate hermiticity in a specific way. This raises the questions: Why does this non-hermiticity arise, and is there an hermitian Hamiltonian for a general curved spacetime having non-neglible time dependence?

Here, we show how to generalize the Hamiltonian of \cite{Parker:1980kw} so that it becomes exactly hermitian without neglecting the time-dependence of the metric. The key is to consider
the Hilbert space structure of the quantum mechanics of the Dirac electron.
We find that the problem with hermiticity that arises when the
metric is varying with time results from a subtle time dependence 
of the basis states $\st{x}$ (i.e., the eigenstates of position).
Once this subtlety is
taken into account, we are able to obtain an expression for the
Hamiltonian of the Dirac fermion that is exactly
Hermitian in a general curved spacetime having an arbitrary space- and time-dependent metric.

The results found in \cite{Parker:1980kw} for the perturbed spectrum of the atom are not affected, but now it
is possible to explore by means of perturbation theory in curved spacetime quantum mechanical effects on bound systems,
such as molecules and atoms, that may result from significant time-dependence of the Riemann tensor.  It would be interesting to determine if such effects could be observed.

\section{Hamiltonian of a Spin-1/2 Particle in a Curved Background}
The Dirac equation in curved spacetime is
\be\el{DiracE} ({\underline \xg}^\mu (x) \nabla_\mu + m )\psi(x) = 0,\ee
where the  ${\underline \xg}^\mu (x)$ matrices are defined by
\be\el{gammabarE} {\underline \xg}^\mu{\underline \xg}^\nu
+{\underline \xg}^\nu{\underline \xg}^\mu = 2g^{\mu\nu}.\ee
The covariant derivative of the spinor $\psi(x)$ is
\be
\el{covaspin} \nabla_\mu \psi(x) \equiv (\p_{\mu} - \Gamma_{\mu})\psi(x),
\ee
where $\Gamma_{\mu}$ is the spinor affine connection.
The spinor covariant derivative of ${\underline \xg}{}_{\nu} (x)$ is
\be \el{derigamma}
 \nabla_\mu \underline {\xg}{}_{\nu} = 
 \p_{\mu} {\underline \xg}{}_{\nu}
 - \Gamma^{\xl}{}_{\mu\nu}{\underline\xg}{}_{\xl}
 -\Gamma_{\mu} {\underline \xg}{}_{\nu}
 +{\underline \xg}{}_{\nu}\Gamma_{\mu} = 0,
\ee
which must vanish so that the covariant derivative of the metric
will be $0$.

A convenient representation of the matrices 
${\underline \xg}^{\mu} (x)$ is
\be {\underline \xg}^{\mu} (x) \equiv {b_a}^{\mu}(x) \xg^a,\ee
where ${b_a}^{\mu}$ is the vierbein (often denoted by
${e_a}^{\mu}$) defined by 
$g^{\mu\nu} = {b_a}^{\mu}{b_b}^{\nu} \eta^{ab}$,
and the $\xg^a$ are the flat spacetime 
gamma-matrices, satisfying
${\xg}^a { \xg}^b
+{ \xg}^b{ \xg}^a = 2\eta^{a b}.$ 
We use the conventions that the metric in Minkowski space is 
$\eta_{ab} = {\rm diag}(-1, 1, 1, 1)$ and therefore
$\xg_0^{\dagger} = -\xg_0$ and
$\xg_i^{\dagger} = \xg_i$. 
The corresponding representation of the spinor affine connection 
$\Gamma_{\mu}$ is
\be
\xG_\mu = -\frac 1 4 \xg_a \xg_b {b^a}_\nu g^{\nu \xl} {b^b}_{\xl;\mu} + i q A_\mu.
\ee
The ``;" here acts on the vierbein as a curved-spacetime vector
\be
{b^b}_{\xl;\mu} \equiv \p_\mu {b^b}_{\xl} - \xG^\rho_{\mu \xl} {b^b}_{\rho}.
\ee
Here, $A_\mu$ is the electromagnetic vector potential. For the atom $A_\mu$ is important, but in considering the Hermiticity of the Hamiltonian we can set $A_\mu = 0$ because it does not contribute to the non-Hermiticity. Therefore, in the following discussion, we will set $A_\mu = 0$.\\ 

It is possible to interpret $\psi(x)$ as the wave function of a spin-1/2 particle moving in curved spacetime.  In Dirac notation \footnote{We will suppress the spinor index.},  it is
$\br x | \psi \ke$. We will take a closer look at that later. The scalar product for the wave function is defined to be \cite{Parker:1980kw},
\be\label{eq:scalarProduct}(\phi, \psi) = -\int d^3 x \sqrt{-g} 
\phi^\dagger(x) \xg^0 {\underline \xg}^0 (x) \psi(x).\ee
\\

It is straightforward to rewrite the Dirac equation \er{DiracE} in the form of a Shr\"odinger equation,
\be\el{Shro} i\frac \p {\p t} \psi(x) = \hat H \psi(x),\ee
where $\hat H$ is given by 
\be\el{Hami} \hat H \equiv - i \iv {g^{00}} {\underline \xg}^0 {\underline \xg}^i \nabla_i + i \xG_0 - i \iv {g^{00}} {\underline \xg}^0 m .\ee
However, as mentioned in \cite{Parker:1980kw},
the Hamiltonian defined in this way is not hermitian when
the metric explicitly depends on the time $t$. One finds that
\goodbreak
\ba 
\el{nonHermi1} 
(\phi, \hat H\psi) - (\hat H \phi, \psi) & = & - \int d^3 x \sqrt{-g} \phi^\dagger \xg^0 {\underline \xg}^0  (-i \iv {(g^{00})} \underline \xg^0 \underline \xg^i \nabla_i + i \xG_0 \nn 
& & - i \iv {(g^{00})} \underline \xg^0 m )\psi + \int d^3 x \sqrt{-g} \Big[{(\nabla_i \phi)}^\dagger i \iv {(g^{00})} {\underline \xg^i}^\dagger {\underline \xg^0}^\dagger) \nn 
& & - \phi^\dagger  i \xG_0^\dagger + \phi^\dagger i \iv {(g^{00})} m {\underline \xg}^{0\dagger} \Big]\xg^0 {\underline \xg}^0 \psi  \nn
& = & \int d^3 x \sqrt{-g} \Big[{(\nabla_i \phi)}^\dagger (i \xg^0 \underline \xg^i )\psi - i\phi^\dagger \xG_0^\dagger\xg^0 {\underline \xg}^0 \psi + i\phi^\dagger\xg^0 \underline \xg^i \nabla_i \psi \nn 
& & - i \phi^\dagger \xg^0 {\underline \xg}^0  \xG_0\psi \Big]\nn
& = & \int d^3 x \sqrt{-g} \Big[-i \phi^\dagger \frac 1 {\sqrt{-g}} \p_i(\sqrt{-g}\xg^0 \underline \xg^i )\psi - i \phi^\dagger \xG_i^\dagger\xg^0 \underline \xg^i \psi \nn 
& & - i\phi^\dagger \xG_0^\dagger\xg^0 {\underline \xg}^0 \psi - i\phi^\dagger\xg^0 \underline \xg^i \xG_i \psi - i \phi^\dagger \xg^0 {\underline \xg}^0  \xG_0\psi \Big]\nn
& = & \int d^3 x \sqrt{-g} \Big[-i \phi^\dagger (\p_\mu+\xG^\nu_{\nu \mu}) (\xg^0 \underline \xg^\mu ) \psi + i \phi^\dagger \frac 1 {\sqrt{-g}} \p_0 (\sqrt{-g}\xg^0 \underline \xg^0 )\psi \nn 
& & + i \phi^\dagger \xg^0 \xG_i \underline \xg^i \psi + i\phi^\dagger \xg^0 \xG_0 {\underline \xg}^0 \psi - i\phi^\dagger\xg^0 \underline \xg^i \xG_i \psi  - i \phi^\dagger \xg^0 {\underline \xg}^0  \xG_0\psi \Big]\nn
& = & \int d^3 x \Big[+ i \phi^\dagger \frac 1 {\sqrt{-g}} \p_0 (\sqrt{-g}\xg^0 \underline \xg^0 )\psi - i \phi^\dagger \xg^0 \nabla_\mu (\underline \xg^\mu ) \psi \Big]\nn
& = & i\int d^3 x \phi^\dagger \xg^0 \frac \p {\p t} \lb \sqrt{-g}  {\underline \xg}^0 \rb \psi.
\ea
In obtaining the 3rd equality we used \er{covaspin}, and to obtain the 4th and 5th equalities we used \er{derigamma}. 
In summary,
\be
\el{nonHermi}
(\phi, \hat H\psi) - (\hat H \phi, \psi) = i\int d^3 x \phi^\dagger \xg^0 \frac \p {\p t} \lb \sqrt{-g} {\underline \xg}^0 \rb \psi.
\ee
The right hand side of \er{nonHermi} is generally nonzero.\\ 

This apparent paradox concerning the non-hermiticity of $\hat H$ in fact comes from the definition of $\psi(x)$. It is defined as $\br \vec x | \psi\ke$ (where $\vec x$ denotes the spatial coordinates).  We must require that
\be
\br \phi | \psi \ke = (\phi, \psi),
\ee
where $(\phi, \psi)$ is the conserved scalar product defined in
Eq.~(\ref{eq:scalarProduct}).  It follows that
the complete bases $\st {\vec x}$ actually satisfies
\be\el{comple} \int d^3x \st {\vec x} \sqrt{-g}\xg^0 {\underline \xg}^0 (x) \br \vec{x}| = 1.\ee
Therefore, when 
$\sqrt{-g}$ depends on time, 
so does $\st {\vec x} \equiv \st {\vec x, t}$. 
As a result,
\be 
\el{Shro1}
i \frac \p {\p t} \br \vec{x},t | \psi \ke \ne i \br \vec{x},t | \lb \frac \p {\p t} \st{\psi}\rb = \br \vec{x},t | {\cal H} |\psi\ke,
\ee
where ${\cal H}$ is the hermitian Hamiltonian in the Schr\"odinger dynamical picture in the abstract Hilbert space. It is the operator that satisfies
\be\el{ShroH} i\frac \p {\p t} \st{\psi} = {\cal H} \st{\psi}.\ee
Note that the left-hand side of \er{Shro1} is what appear on the 
left of \er{Shro}. 
In other word, the $\hat H$ (defined in \er{Hami}), which is on the 
right-hand side of \er{Shro}, is not quite the Hamiltonian in the
Schr{\"o}dinger or configuration-space representation when the
metric depends on $t$.\\ 

Let us find the matrix elements
$\br{\vec x, t}| \mathcal{H} \st{\vec x', t}$.
One can show from \er{comple} that
\be\frac \p {\p t} \st{\vec x, t} = 
-\frac 1 2 \st{\vec x, t}  \frac \p {\p t} \lb \sqrt{-g} \xg^0 {\underline \xg}^0 (x)\rb {(\sqrt{-g}\xg^0 {\underline \xg}^0 (x))}^{-1}.\ee
Taking the conjugate, we have 
\ba\frac \p {\p t} \br\vec x, t| & = &
-\frac 1 2  {{(\sqrt{-g}\xg^0 {\underline \xg}^0 (x))}^\dagger}^{-1} \frac \p {\p t} {\lb \sqrt{-g} \xg^0 {\underline \xg}^0 (x)\rb}^\dagger \br\vec x, t|\nn
& = & -\frac 1 2  {(\sqrt{-g}\xg^0 {\underline \xg}^0 (x))}^{-1} \frac \p {\p t} \lb \sqrt{-g}\xg^0 {\underline \xg}^0 (x) \rb \br\vec x, t|.
\ea
Here we used the fact that $\xg^0 {\underline \xg}^0 (x)$ is Hermitian. It is then easy to see that the completeness relation of \er{comple} is independent of time.
 We also have
\be{(\xg^0 {\underline \xg}^0 (x))}^{-1} = \frac {{\underline \xg}^0 (x)\xg^0} {-g^{00}}.\ee
So the Hamiltonian $H$ satisfying the condition
$(\psi, H \phi) = (H \psi, \phi)$
is
\ba
\el{Hami1} H & \equiv & -i \frac 1 2  \frac {{\underline \xg}^0 (x)\xg^0} {g^{00}\sqrt{-g}} \frac \p {\p t} \lb \sqrt{-g} \xg^0 {\underline \xg}^0 (x)\rb - i \iv {g^{00}} {\underline \xg}^0 {\underline \xg}^i \nabla_i + i \xG_0 - i \iv {g^{00}} {\underline \xg}^0 m \nn 
&  = & -i \frac 1 2  \frac {{\underline \xg}^0 (x)\xg^0} {g^{00}\sqrt{-g}} \frac \p {\p t} \lb \sqrt{-g} \xg^0 {\underline \xg}^0 (x)\rb +\hat H.\ea
It thus follows that $H$ is Hermitian (with the use of \er{nonHermi})
for a general metric.
The matrix elements of ${\cal H}$ satisfy the relation
\be\el{xrep}\br{\vec x, t}| \mathcal{H} \st{\vec x', t} 
=  H  \delta(\vec{x} - \vec{x}')(\sqrt{-g})^{-1}, \ee
with $H$ being the operator given in Eq.~(\ref{Hami1}).\\

If we are dealing with a one-electron atom, the spinor
affine connections will contain the vector potential of the
electromagnetic field, but the derivation is unchanged. Let us
consider the effect of the time dependence of the 
Riemann tensor on the spectrum of the atom.
The Hamiltonian $H$ of Eq.~(\ref{Hami1}) reduces to $\hat H$ 
if the time dependence
of the metric can be neglected, as was the case 
in \cite{Parker:1980kw}.  For rapidly changing gravitational fields
the additional term is needed to enforce Hermiticity.

In the Fermi normal coordinates along the geodesic of
a bound system such as an atom, the difference, $H-\hat H$,
given by \er{Hami1} must vanish on the geodesic because
it involves only the first time-derivative of the curvature. 
Furthermore, for
small distances from the geodesic, this difference, $H-\hat H$, is not vanishing, but it is of higher
order in $\frac {a_0}{r}$ (where $a_0$ is the atomic size and $r$ is a
characteristic length or time scale of the background spacetime) compared to the other terms in $H$. This can be seen by dimensional analysis from the Hamiltonian $\hat H$ that is
given in Fermi normal coordinates in \cite{Parker:1980kw}.
Therefore, when $\frac {a_0}{r} \ll 1$ the difference between
$H$ and $\hat H$ can be neglected.
Similarly, when $\frac {a_0}{r} \ll 1$, 
it is also possible to use $\hat H$  with
time-dependent perturbation theory to calculate transition
rates induced by the Riemann tensor along the path of the atom.

\section{Conclusion}
We have revisited the quantum mechanics of a one-electron atom 
in an arbitrary curved background.  We addressed the following problem.
The operator $\hat H$, which appears on the right-hand side of \er{Shro}, $i\partial_t\psi(x) = \hat H\psi(x)$, is not
hermitian with respect to the curved-spacetime scalar product.
But \er{Shro} was obtained directly from the Dirac equation in curved spacetime, so why is $\hat H$ not Hermitian? We resolved this apparent paradox in the following way.
We started from the fundamental Schr{\"o}dinger equation
$ i\partial_t \st{\psi} = {\cal H} \st{\psi}$ of \er{ShroH},
where the operator ${\cal H}$ is Hermitian.
From the completeness relation, \er{comple}, we showed that the eigenstates of position that span the Hilbert space must depend on time as well as spatial position: $\st{\vec{x}, t}$.
The wave function $\psi(x)$ is defined as $\br \vec{x}, t \st{\psi}$.
By applying $\br{\vec{x}, t}|$ from the left to Eq.~(\ref{ShroH}), we found that the position-space representation of ${\cal H}$ is given by \er{xrep}. 
The differential operator
$H$ that appears in this representation is Hermitian with respect
to the curved-spacetime scalar product. 
However, the time derivative of $\br{\vec{x}, t}|$ in the wave function
gives an additional terms in $H$ that does not appear in 
$\hat{H}$.  Thus, we see why \er{Shro} is correct, but does not
involve the Hermitian operator $H$.  We have also discussed
the circumstances in which $\hat H$ is effectively hermitian and can
be used to do perturbation theory to find shifts in energy levels
and transition rates.

\appendix
\section{FRW Metric}

As a simple example, consider a general Robertson-Walker metric
given by
\[ds^2 = - dt^2 + a^2 (t) (dx^2 + dy^2 + dz^2).\]
For this metric, one finds \cite{Parker:1971pt} that
\begin{equation}
{\underline \xg}^0 = \xg^0,\quad {\underline \xg}^i = a(t)^{-1}\xg^i,
\quad \xG_0 = 0,\quad \xG_i = \frac 1 2 \dot a (t) \xg^0 \xg^i .
\end{equation}
and the curved spacetime scalar product looks like that in
Minkowski space, except for a factor of $a(t)^3$.
In the present case, the $\xG^0$ term in $H$ does not appear, and
the term $- i \iv {g^{00}} {\underline \xg}^0 m$ is the same as in flat space time. Therefore, the only part that may be non-Hermitian is the $- i \iv {g^{00}} {\underline \xg}^0 {\underline \xg}^i \xG_i$ that comes from the covariant derivative $- i \iv {g^{00}} {\underline \xg}^0 {\underline \xg}^i \nabla_i$. The other term from the covariant derivative, $- i \iv {g^{00}} {\underline \xg}^0 {\underline \xg}^i \p_i$ is Hermitian because $\p_i$ commutes with ${\underline \xg}^i$.
So we have
\ba (\phi, \hat H\psi) - (\hat H \phi, \psi) & = & - \int d^3 x \sqrt{-g} \Big[ \phi^\dagger \xg^0 {\underline \xg}^0 (x) (i \iv {(g^{00})} \underline \xg^0 \underline \xg^i (x)\xG_i \psi - \phi^\dagger (i \iv {(g^{00})}\xG_i^\dagger {\underline \xg^i (x)}^\dagger {\underline \xg^0}^\dagger  \xg^0 {\underline \xg}^0 (x) \psi\Big].\nn
& = & \int d^3 x \sqrt{-g} \Big[ \phi^\dagger \xg^0 (i\xG_i \underline \xg^i (x) -i \underline \xg^i (x)\xG_i)\psi\Big],
\ea 
which reduces to 
\ba (\phi, \hat H\psi) - (\hat H \phi, \psi) & = & -3 i \int d^3 x \sqrt{-g}\phi^\dagger \frac {\dot a (t)} {a(t)} \psi. \nn
& = & i\int d^3 x \phi^\dagger \xg^0 \frac \p {\p t} \lb \sqrt{-g}  {\underline \xg}^0 (x)\rb \psi.
\ea
So we can see $\hat H$ is indeed non-Hermitian and the violation of hermiticity is given by \er{nonHermi}. Therefore the operator $H$ 
defined in \er{Hami1} is hermitian in this spacetime. This agrees with the result in \cite{Audretsch:1978zw}, which deals with a cosmological metric.

\goodbreak

\end{document}